\documentclass
[pra,aps,preprint,superscriptaddress,twocolumn,showpacs,10pt]{revtex4}%
\usepackage{amsmath}
\usepackage{amsfonts}
\usepackage{amssymb}
\usepackage{graphicx}
\usepackage{threeparttable}
\usepackage[colorlinks, citecolor=blue]{hyperref}%
\providecommand{\U}[1]{\protect \rule{.1in}{.1in}}
\providecommand{\U}[1]{\protect \rule{.1in}{.1in}}

\begin{document}

\title{Wave-function Visualization of Core-induced Interaction of Non-hydrogenic
Rydberg Atom in Electric Field}
\author{W. Gao}
\affiliation{State Key Laboratory of Magnetic Resonance and Atomic and Molecular Physics,
Wuhan Institute of Physics and Mathematics, Chinese Academy of Sciences, Wuhan
430071, People's Republic of China}
\affiliation{University of Chinese Academy of Sciences, Beijing 100049, People's Republic
of China}
\author{M. Deng}
\affiliation{State Key Laboratory of Low-Dimensional Quantum Physics, Department of
Physics, Tsinghua University, Beijing 100084, China}
\author{H. Cheng}
\affiliation{State Key Laboratory of Magnetic Resonance and Atomic and Molecular Physics,
Wuhan Institute of Physics and Mathematics, Chinese Academy of Sciences, Wuhan
430071, People's Republic of China}
\affiliation{University of Chinese Academy of Sciences, Beijing 100049, People's Republic
of China}
\author{S. S. Zhang}
\affiliation{State Key Laboratory of Magnetic Resonance and Atomic and Molecular Physics,
Wuhan Institute of Physics and Mathematics, Chinese Academy of Sciences, Wuhan
430071, People's Republic of China}
\affiliation{University of Chinese Academy of Sciences, Beijing 100049, People's Republic
of China}
\author{H. P. Liu\footnote{E-mail: liuhongping@wipm.ac.cn}}
\affiliation{State Key Laboratory of Magnetic Resonance and Atomic and Molecular Physics,
Wuhan Institute of Physics and Mathematics, Chinese Academy of Sciences, Wuhan
430071, People's Republic of China}
\affiliation{University of Chinese Academy of Sciences, Beijing 100049, People's Republic
of China}

\begin{abstract}
We have investigated the wave-function feature of  Rydberg sodium in
a uniform electric field and found that the core-induced interaction of
non-hydrogenic atom in electric field can be directly visualized in the
wave-function. As is well known, the hydrogen atom in electric field can be
separated in parabolic coordinates $(\eta, \xi)$, whose eigen-function can
show a clear pattern towards negative and positive directions corresponding to
the so-called red and blue states without ambiguity, respectively. It can be served as a
complete orthogonal basis set to study the core-induced interaction of
non-hydrogenic atom in electric field. Owing to complete different patterns of
the probability distribution for red and blue states, the interaction can
be
visualized in the wave-function directly via superposition. Moreover, the constructive and
destructive interferences between red and blue states are also observed in the
wave-function, explicitly explaining the experimental measurement for the
spectral  oscillator strength.

\end{abstract}
\date{\today}
\maketitle
\volumeyear{ }
\volumenumber{ }
\issuenumber{ }
\eid{ }
\received[Received text]{}

\revised[Revised text]{}

\accepted[Accepted text]{}

\published[Published text]{}

\startpage{1}
\endpage{ }


\section{Introduction}

Ever since the pioneering work on the stark spectra of alkali-metal atoms made
by Zimmerman et al in \cite{104}, the effect of electric field on
non-hydrogenic atoms in Rydberg states has been investigated extensively both
theoretically and experimentally \cite{283,560,311,1173}. In zero field, the
difference between hydrogen and non-hydrogenic atoms is reflected by the large
quantum defect values embodying core effect for low angular momentum states.
When the non-hydrogenic atom is placed in a uniform electric field, this impure Coulomb potential couples
levels from different principal quantum number $n$, resulting in different ionization mechanism \cite{61,371}. Several
works have been performed to explore the non-hydrogenic behaviour.

Stark map, plotting the level splitting of Rydberg atoms against the
increasing electric field strength, is a very direct and popular way to reveal
the anticrossing behaviour of non-hydrogenic atoms in electric field. Many
theoretical approaches have been developed to deal with Rydberg atoms in
electric field. For instance, explicit expressions of the perturbation theory
up to seventh order was presented by Silverstone in \cite{1099}, which is very
accurate and effective in obtaining the hydrogenic stark levels. Derived from
the Gutzwiller trace formula, the closed-orbit theory was used to calculate
photoabsorption cross sections of hydrogen and sodium in a strong electric
field \cite{178,179}. Theoretical work with WKB theory was also conducted in
Ref. \cite{283}, which basically reproduced the photoionization cross sections of
the excited $3^{2}\mathrm{{P_{3/2}}}$ state of sodium. When the influence of
external fields strongly dominates over the core potential, quantum
calculations based on finite basis expansion are usually employed for more
accurate results. Diverse basis sets have been adopted, including B-splines basis \cite{1257}, Sturmian basis \cite{511}, potential
model basis \cite{454} and quantum defect orbital \cite{85}.

Among these theoretical methods, the most interesting is the closed-orbit
theory closely connecting the photoabsorption spectra and classical orbits,
which provides an intuitive and visual physical picture for the non-hydrogenic
behaviour. The so-called scaled-energy spectroscopy interpreted the
non-hydrogenic lithium stark effect in the term of combined orbits indicating
core scattering from one closed orbit to another one \cite{311}. West \emph{et
al.} also calculated and animated the field-induced dynamics with classical
atomic models for circular states \cite{1173}. Their numerical simulations of
a classical electron trajectory showed stark oscillations in a dc electric
field. For a pure Coulomb potential, the electron backscatters in the linear
obit and nearly retraces the original trajectory in the opposite direction,
but for the trajectory in the sodium model potential, a precession of the
highly eccentric orbit is found.

More recently, motivated by theoretical predictions of `photoionization
microscopy' \cite{1261,1268}, Stodolna \emph{et al.} made the first direct
visual observation of the nodal structure of stark states for hydrogen
\cite{896} with the help of photoelectron imaging \cite{1259} and an
electrostatic zoom lens \cite{1270}. The observed interference pattern clearly
revealed that the number of dark fringes corresponds to the parabolic number
$n_{1}$ of the excited quasibound atomic state. It was also visualized
previously in the frameworks of both a semiclassical open-orbit theory and a
quantum-mechanical theory developed by Zhao and Delos in \cite{362,1087}.
Based upon Harmin's semiclassical theory \cite{403,283}, Robicheaux and Shaw
calculated the wave-packet dynamics of a Rydberg electron in a strong electric
field for rubidium \cite{1271}. Different from the WKB approximation, the
coupled-channel theory was also developed to stimulate experimental
observations for non-hydrogenic atoms \cite{1089}. Subsequently, experimental
results performed on helium \cite{1085} indirectly visualized the coupling
between red and blue stark states with the aid of the interference narrowing
\cite{33}.

The visualization of the wavefunction as well as the coupling between red and
blue states is based on a unique feature of hydrogen atom that the Hamiltonian
in a static field is separable in terms of the parabolic coordinates $\eta,
\xi$ ($\eta= r - z$, $\xi= r + z $) and can be strictly solved. Since the
parabolic separation for hydrogen in electric field can supply a set of
`orthogonal' basis $\{$red states, blue states$\}$, can we reveal the
non-hydrogenic effects directly from the eigen-wavefunctions of nonhydrogenic
atom in electric fields? In this paper, we will investigate whether the
rigorous hydrogenic parabolic states can be served as a set of basis in deeper
understanding and visualizing complex dynamics of non-hydrogenic atom in
electric field. For this purpose, taking the effect of the atomic core on Rydberg electron into account, we computed the oscillator strength and
electron probability distributions for hydrogen and sodium. By comparing their
electron probability distributions directly, we show that wave-function can be
intuitive in understanding the coupling behaviour of non-hydrogenic atom in
static electric fields on benefit of the `orthogonality' of red states and
blue states. Moreover, the irregular oscillator strength distribution for
sodium is analyzed in the perspective of electric-field-induced interference
between hydrogenic parabolic states. The analysis is also compared with our
experimental spectrum of sodium in electric field of $840 \mathrm{{V/cm}}$
below and above the saddle point $E_{c}$.

\section{Theoretical Calculation}

Since the exact quantum defect theory (EQDT) has been detailed previously
\cite{215,281}, only a brief summary is given here. The sodium in a
uniform electric field oriented along the $z$-axis can be described by an
exact non-relativistic Hamiltonian, hereafter in atomic units, taking the
form
\begin{equation}
H = \frac{p^{2}}{2} + V^{c}(r) + Fz, \label{eq1}%
\end{equation}
where $V^{c}(r)$ is the Coulomb potential including the effects of
core-induced electron screening, leading to non-separable of Hamiltonian in
parabolic coordinates. Instead of using the model potential \cite{454} or the
R-matrix method \cite{32} including the quantum defects implicitly, we employ
an equivalent form for the central field potential \cite{911,551}:
\begin{equation}
V(r)=\frac{\lambda(\lambda+1)-l(l+1)}{2r^{2}}-\frac{1}{r}, \label{eq2}%
\end{equation}
where $\lambda=l-\delta+Int(\delta)$ and the quantum defects are explicitly
enclosed. Here $Int(\delta)$ is the rounded nearest-integer value of the
quantum defect. A reduced quantum defect $\delta^{\prime}=\delta-Int(\delta)$
is employed here to assess the real contribution of the quantum defect for a
given angular momentum channel. We will directly use $\delta$ to stand for
$\delta^{\prime}$ for abbreviation.

Solving the time-independent Schr\"{o}dinger equation
\begin{equation}%
\begin{split}
H\psi= E\psi,\label{eq3}%
\end{split}
\end{equation}
we can compute the energy levels and their wavefunctions, from which
transition probabilities are calculated.

The radial part of the wavefunction is expanded in the B-spline basis and the
angular part in the truncated associated Legendre function basis,
\begin{equation}%
\begin{split}
\psi(r,\theta) = \sum \limits_{n = 0}^{N - 1} \sum \limits_{l = \left|  m
\right|  }^{{l_{\max}}}C_{nl} \frac{^{ }{B_{n}^{k}(r)}{}}{r}{P_{l}^{\left|  m
\right|  }}{(\theta)},\label{eq4}%
\end{split}
\end{equation}
where $P_{l}^{^{\left|  m \right|  }}(\theta)$ is the normalized associated
Legendre function and $B_{n}^{k}(r)$ the $k$ order B-spline function defined
in Refs.\cite{338,416}.

Substituting the Hamiltonian expressed in Eq.(\ref{eq1}) and the wavefunction
in Eq.(\ref{eq4}) into Eq.(\ref{eq3}), the Schr\"{o}dinger equation is
transformed into a general eigenvalue problem,
\begin{equation}
HC = ESC, \label{eq5}%
\end{equation}
where $E$ and $C$ represent eigenvalues and their corresponding eigenvectors,
respectively. $H$ is the matrix form of the Hamiltonian and $S$  the
overlap matrix. Accurate matrix elements are obtained efficiently through the
Gauss-Legendre quadrature scheme. A Lanczos algorithm \cite{1256} for the
general eigenproblem applied to the matrix equation can give the eigenvalue
$E$ and eigenvector $C$, which makes the calculation very effective. The
oscillator strength for the one-photon dipole transition from the ground
state $|i\rangle$ to the final state $|f\rangle$ can be expressed as
\begin{equation}%
\begin{split}
{c_{i \to f}} = {\left|  {\left \langle {i\left|  d \right|  f} \right \rangle }
\right|  ^{2}},\label{eq6}%
\end{split}
\end{equation}
where $d$ is the electric dipole operator.

\section{Results and Discussions}

To study the visual feature of the wavefunction of atom with core-induced
interaction in electric field, we have to choose an appropriate energy range
within which the spectrum can be well resolved experimentally, and at the
time, theoretically, it should be simple enough for the wavefunction analysis.

\begin{figure}[ptb]
\centering
\includegraphics[width=3.3in]{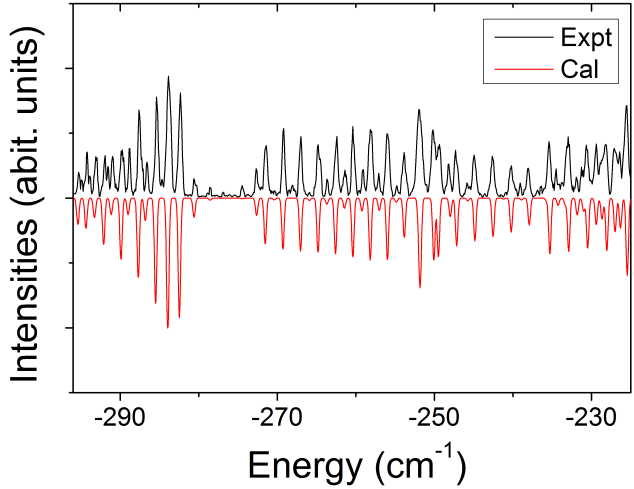}\caption{(Color online) Comparison of
the experimental (upper panel) and theoretical (lower panel) stark spectra for
Rydberg sodium in the electric field $F = 840$ V/cm with $\pi$-polarized laser
irradiation.}%
\label{fig1}%
\end{figure}

We experimentally measure the stark spectrum of Rydberg sodium in electric
field of $840$ V/cm in the energy range $-290$ cm$^{-1} <E<-230$ cm$^{-1}$ and
carry out a theoretical calculation with the present EQDT method. They are
shown in Fig.\ref{fig1}. In the experiment, the laser is linearly polarized
parallel to the electric field $F$, thus the $\Delta m = 0$ transition occurs.
In the calculation, we adopt the reduced quantum-defect values from
Ref. \cite{104}, taking $\delta_{s} = 0.347$, $\delta_{p} = -0.146$ and
$\delta_{d} = 0.014$. Quantum defect values for all higher angular momenta are
negligibly small and set as zero. The theoretical oscillator strength is
convoluted with a Gaussian profile of $0.18$ $\mathrm{{cm^{-1}}}$ to give
a spectrum comparable with experimental observation. The Gaussian broadening
includes  the Doppler broadening and the laser linewidth. To facilitate a direct
comparison, the experimental observation and the convoluted spectrum are shown
in mirror fashion. It's obvious that for the current electric field strength
and energy range, the experimental resonances are well resolved to be
comparable to calculation. Although trivial differences in relative
intensities for several resonances exist due to the inhomogeneity in the
electric field and instability of the laser power, the agreement between
theoretical calculation and experimental observation is rather satisfactory
both for the positions and intensities under present condition. This serves as
a strong evidence of the validity of EQDT method and supplies a reliable
groundwork for the analysis of  the oscillator strength re-distribution and
wavefunction feature.

\begin{figure}[ptb]
\centering
\includegraphics[width=3.3in]{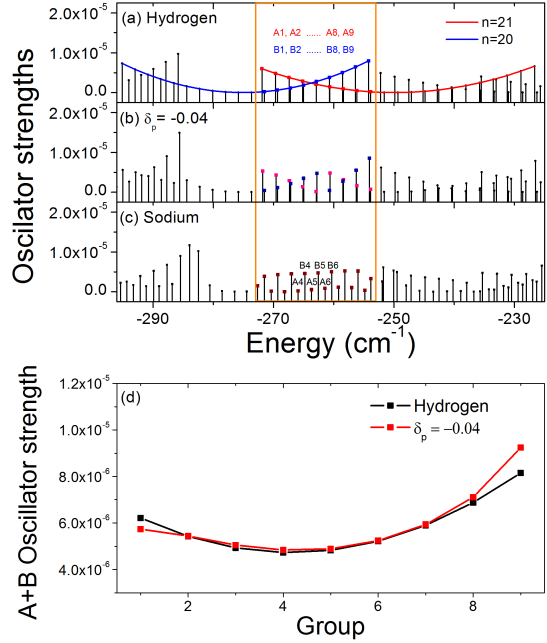}\caption{(Color online) Theoretical
calculation of photoexcitation spectrum from the ground state to stark
manifolds $n = 19$ to $23$ with laser polarized parallel to the electric field
$F = 840$ V/cm for (a) hydrogen atom, (b) artificial atom with $\delta
_{p}=-0.04$ and (c) sodium atom, and the oscillator strength sums of the
interacting spectral peaks (d). The complete stark manifold of $n = 20$ (blue
dots + line) and $n= 21$ (red dots + line) of hydrogen in (a) are specially
marked out to trace the symmetric envelop and make the crossing characteristic
clear. Note that the spectral lines in crossing area are numbered to
facilitate the analysis. For the closely interacting pair of peaks for the
artificial atom, their oscillator strength sums are kept constant for the
first order approximation, compared with hydrogen atom.}%
\label{fig2}%
\end{figure}

In Fig.\ref{fig1}, we can also notice that the spectral lines, for example, in
the energy range $-270\sim-250$ cm$^{-1}$, are accompanied with some lines
with weak intensities. These weak lines are due to the core-induced
interaction for sodium in electric field. As a result, the spectral
line intensity re-distribution is an indicator for the non-Coulombic potential interaction. To
make it clear, we calculate a series of oscillator strengths for three
different atomic systems, from simple to complex, including hydrogen,
artificial non-hydrogenic atom and sodium. Specifically, the artificial
non-hydrogenic atom has only one quantum defect channel $\delta_{p} = -0.04$,
more complex than hydrogen atom but much simpler than the real sodium atom.
Sodium atom has non-ignorable quantum defects for $s$, $p$ and $d$ channels.
We calculate the oscillator strengthes for these three types of atoms in the
energy range corresponding to that in Fig.\ref{fig1}, which
is shown in Fig.\ref{fig2}. The energy level positions are listed in
table \ref{tab1}, where we can see that our calculations for hydrogen atom by
the EQDT method are exactly the same as that by the seventh-order perturbation
theory \cite{1099} up to three digital number. Our calculation for sodium by
the EQDT method agree with the experimental observation within the Gaussian
broadening of $0.18$ cm$^{-1}$.

\begin{table*}[ptb]
\centering
\par
{\small \begin{threeparttable}
\caption{The energy level positions (in $\rm{cm^{-1}}$) of the Stark resonances of hydrogen and sodium ($F = 840$ V/cm, $m=0$) within $-270$
cm$^{-1}$ $\sim$ $-250$ cm$^{-1}$ shown in Fig.\ref{fig2}.
}
\begin{tabular}{ccc|ccc}
\toprule
\quad & hydrogen&\quad& \quad& sodium \quad \\
\hline
State:$(n_1,n_2,m)$ &\quad Theory$^a$  &\quad Theory$^b$  &\quad  State &\quad Theory$^c$  &\quad Experiment$^d$ \\
\hline
(19,0,0) & -254.228 & -254.228 & A1 & -272.628 & -272.674\\
(18,1,0) & -256.393 & -256.393 & B1 & -271.534 & -271.403\\
(17,2,0) & -258.558 & -258.558 & A2 & -270.390 & -270.300\\
(16,3,0) & -260.721 & -260.721 & B2 & -269.276 & -269.198\\
(15,4,0) & -262.882 & -262.882 & A3 & -268.159 & -268.011\\
(14,5,0) & -265.042 & -265.042 & B3 & -267.038 & -266.993\\
(13,6,0) & -267.201 & -267.201 & A4 & -265.922 & -265.891\\
(12,7,0) & -269.359 & -269.359 & B4 & -264.811 & -264.788\\
(11,8,0) & -271.515 & -271.515 & A5 & -263.684 & -263.686\\
(0,20,0) & -271.912 & -271.912 & B5 & -262.588 & -262.413\\
(1,19,0) & -269.666 & -269.666 & A6 & -261.447 & -261.395\\
(2,18,0) & -267.419 & -267.419 & B6 & -260.373 & -260.377\\
(3,17,0) & -265.170 & -265.170 & A7 & -259.220 & -259.105\\
(4,16,0) & -262.919 & -262.919 & B7 & -258.164 & -258.172\\
(5,15,0) & -260.666 & -260.666 & A8 & -257.015 & -257.069\\
(6,14,0) & -258.412 & -258.412 & B8 & -255.972 & -256.051\\
(7,13,0) & -256.156 & -256.156 & A9 & -254.839 & -254.778\\
(8,12,0) & -253.898 & -253.898 & B9 & -253.828 & -253.844\\
\hline
\end{tabular}\label{tab1}
\small
\begin{tablenotes}
\item[a] Calculation for hydrogen by the seventh-order perturbation theory \cite{1099};
\item[b] Calculation for hydrogen by the present EQDT method, exactly
the same as that by the above perturbation theory;
\item[c] Calculation for sodium by the present EQDT method;
\item[d] Experimental observation for sodium (laser linewidth is about
$0.09$ cm$^{-1}$ and the total Gaussian broadening is $0.18$ cm$^{-1}$).
\end{tablenotes}
\end{threeparttable}
 }\end{table*}

As shown in Fig.\ref{fig2}(a), for hydrogen atom, the oscillator strength
distribution shows high regularity. For the stark manifold belonging to a
specified principal quantum number $n$, there are two energy levels
corresponding to the same $|n_{1}-n_{2}|$ value, but they have the same
oscillator strength. In addition, for a specified $n$, the oscillator
strengths of a parabolic state will decrease as the value of $|n_{1}-n_{2}|$
decreases and reach minimum when $|n_{1}-n_{2}|$ reaches the minimum.
Kondratovich and Delos have derived a simple semiclassical formula to
interpret the symmetric envelop of a complete stark manifold for the
hydrogenic $m = 0$ case \cite{1192}. They considered that the oscillator
strength of a parabolic state is proportional to the absolute square of the
quantum angular function of the outgoing waves. The most important property of
hydrogen atom in electric field is that energy levels belonging to different
principal quantum number $n$ can cross each other without interaction. This
unique feature is attributed to the SO(4) symmetry of the Coulomb potential of
hydrogen \cite{1273}. Any deviation from the pure Coulomb potential will break this
symmetry and cause the coupling between two hydrogenic energy levels,
resulting in irregularity of oscillator strength distribution of
non-hydrogenic atoms. This irregularity can be seen from the calculated
oscillator strengths for artificial atom and sodium atom, as shown in Fig.\ref{fig2}(b) and (c), respectively. For the artificial atom, there is only
one channel with quantum defect so that it behaves closer to a hydrogen atom,
providing us a theoretical model to investigate the core-induced interaction
by spectral splitting and oscillator strength.

Regarding two energy levels close to each other as a group, the sums of the
two
oscillator strengths for hydrogen and artificial atom are presented in
Fig.\ref{fig2}(d). The oscillator strengths summed over two coupling
hydrogenic parabolic states are almost conserved, specially for the groups
with energy levels close each other. Some discrepancies exist for group 1 and
9, which shows more energy levels take part in the interaction at these
points. This
oscillator strength conservation provides a way to study the core-induced
interaction between red and blue states. For the real atom, sodium, the
interactions are so strong that the group classification is tricky, as shown
in Fig.\ref{fig2}(c). In this case, more than two energy levels participate in
the interaction at any given energy range within  $-270\sim$ $-250$ cm$^{-1}$ and it is difficult to group the spectral lines.

\begin{figure*}[ptb]
\centering
\includegraphics[width=5in]{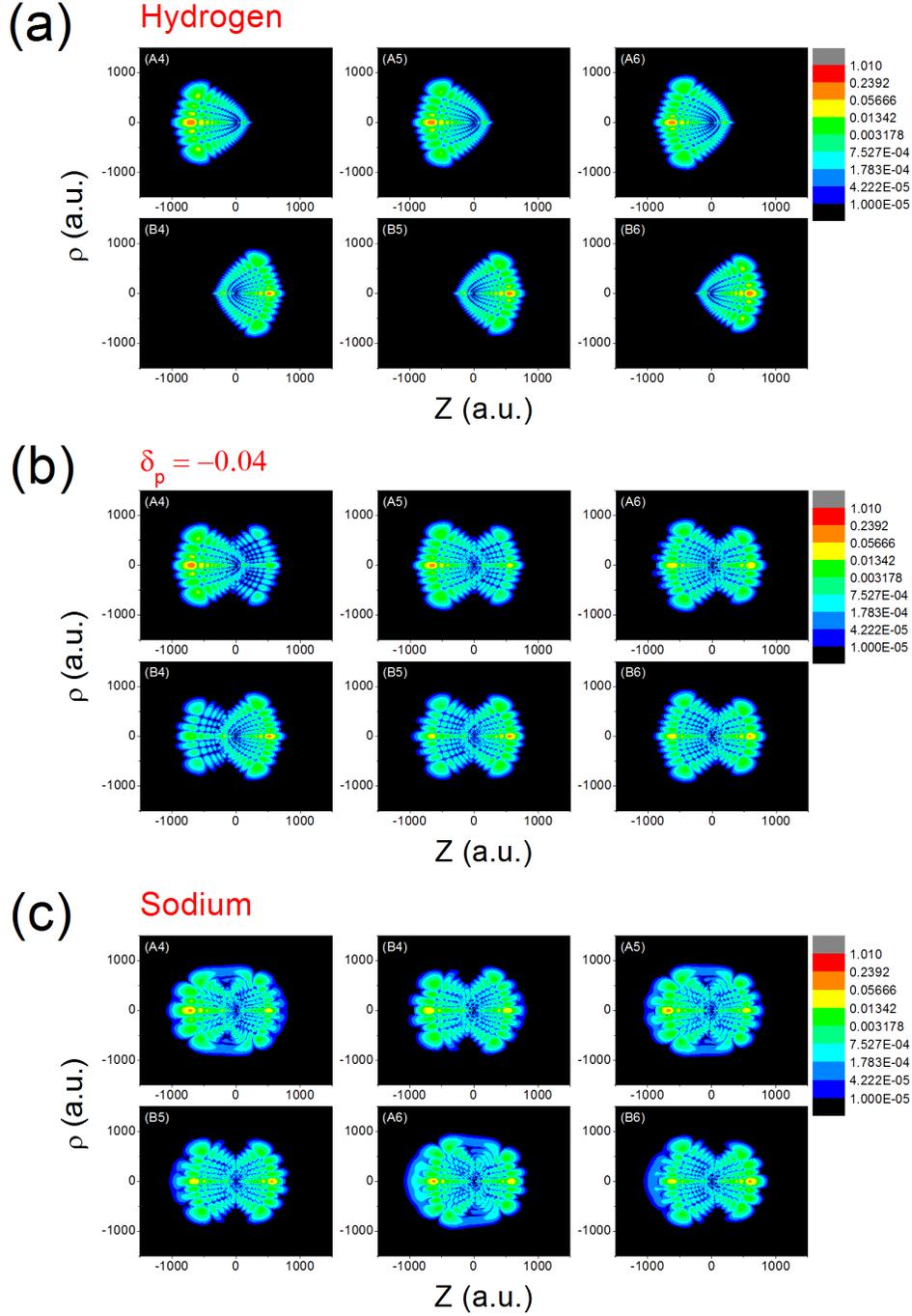}\caption{(Color online) Contour plot of
electron probability distribution $|\Psi(\rho,z)|^{2}$ for the states marked
in Fig.\ref{fig2} for hydrogen (a), artificial atom (b) and sodium (c).}%
\label{fig3}%
\end{figure*}

Rather than studying the interaction in the point of view of energy level, we
can visualize it on the wavefunction directly, which is shown in
Fig.\ref{fig3}. For hydrogen atom shown in Fig.\ref{fig3}(a), the electron
probability distribution $|\Psi(\rho,z)|^{2}$ of a parabolic state is
localized in area of $z > 0$ or $z < 0$, which depends on whether the state is a
blue-shifted state ($n_{1}>n_{2}$) or a red-shifted state ($n_{1}<n_{2}$).
This unique feature benefits from the dynamics symmetry of hydrogen atom.
There is no combined distribution composed of red and blue states \cite{896}. Unlike
hydrogen, however, the electron probability distribution $|\Psi(\rho,z)|^{2}$
for artificial atom and sodium loses this feature \cite{1085}. It looks like a
superposition of two hydrogenic states at first glance, which clearly shows
the core-induced coupling effect in the electron probability distributions. It
is shown in Fig. \ref{fig3}(b-c). For example, the probability distribution of
both A4 or B4 states of artificial atom can be regarded as a superposition of
A4 and B4 states of hydrogen. This point is proved by the numbers of parabola
facing towards the negative and positive $z$-axis. Obviously, for different
state-coupling, the combining weight coefficients are different. Specially,
for the coupling scheme A4 shown in Fig.\ref{fig3}(b), we can see from the
contrast of the probability distribution that the red state is perturbed by
the blue state slightly. As its counterpart, seen from B4, the blue state is
perturbed by the red state in a small extent. While for sodium, the
probability distribution of both A4 or B4 states is not as clear as that for
artificial atom shown in Fig.\ref{fig3}(b) and it can be viewed as a
superposition of more than two red and blue states of hydrogen, which is
observable in Fig.\ref{fig3}(c).


\begin{figure}[ptb]
\centering
\includegraphics[width=3.3in]{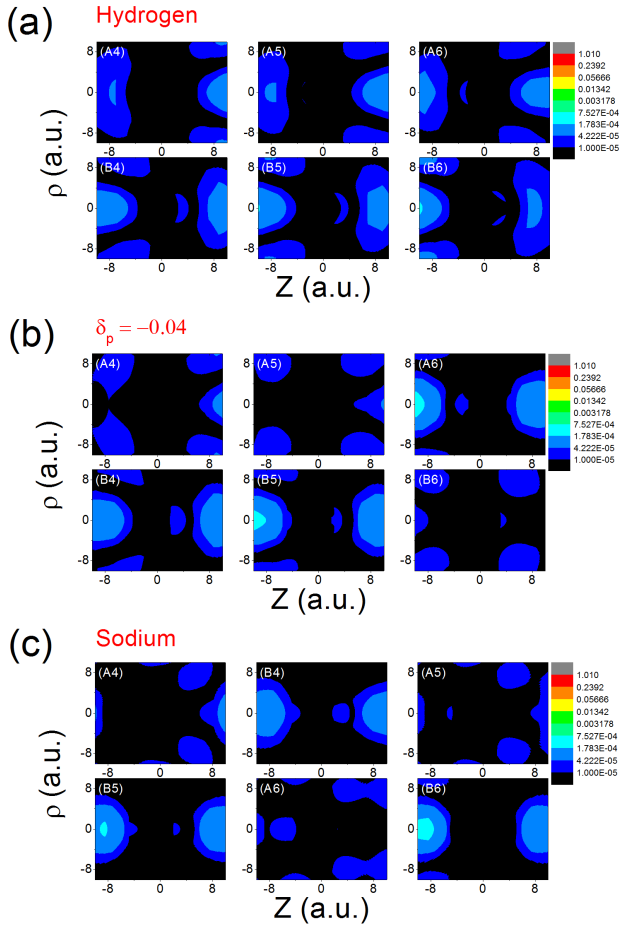}\caption{(Color online) Magnification
of Fig.\ref{fig3} for better evaluation of the constructive and destructive
interference of wavefunction close to the atomic core. For the $\pi$ laser
irradiation, the initial state
has a spindle shape, based on which we can draw a conclusion that the transitions
to the states in (a) can have considerable amplitudes since their probability
distributions have the
same shape. For the non-hydrogenic atoms shown in (b) and (c), however, some
states, for example, A4, A5 and B6 in (b) and A4, A5 and A6 in (c), lose the spindle-like distribution, resulting in weak transitions. }%
\label{fig4}%
\end{figure}

Since the oscillator strength is proportional to the absolute square of the
overlap of the initial state $\langle i|d$ and the final state $|f\rangle$
according to the formula given in Eq.\ref{eq6}, we should be able to observe
the constructive and destructive interference between red and blue states for
the core-induced coupling. As it is known, the ground state of sodium is
$3s^{1}$, having a spherical symmetry. The dipole moment term in Eq.\ref{eq6}
is proportional to $\mathrm{{cos \theta}}$, $d\sim \mathrm{{cos \theta}}$,
resulting in an initial state with spindle shape. On the other hand, the
wavefunction of the ground state is confined in a small radial range, and the
final oscillator strength is dependent on the behavior of the final state wavefunction in
the same range, where we can also observe the constructive and destructive
interference between the red and blue states.

We magnified Fig.\ref{fig3} into Fig.\ref{fig4} to observe the probability
distribution close to the nuclear core, within radius less than $10$ a.u.. In
Fig.\ref{fig4}(a), we can see that the probabilities, either for red states A4
to A6 or blue states B4 to B6, have abundant densities along the $z$-axis,
which sufficiently overlap with the initial state $\langle i|d$ with spindle
shape. All the peaks have close oscillator strengths, in accordance with the
given oscillator strengthes in Fig.\ref{fig2}(a). While for the artificial
atom shown in Fig.\ref{fig4}(b), only the states of A6, B4 and B5  keep
the spindle shape, agreeing with that shown in Fig.\ref{fig2}(b). It is caused by
the constructive interference between the red and blue states. For the left
states A4, A5 and B6, the oscillator strengthes are reduced greatly for the
destructive interference. As a complex case, for the real atom sodium shown in
Fig.\ref{fig4}(c), the states B4, B5 and B6 own the spindle shape, giving
considerable oscillator strengthes, quite consistent with the calculations in
Fig.\ref{fig2}(c) as well, even for the interactions looking rough and tumble.
Anyway, the probability distribution can provide a deeper explanation for the
transition intensity based on the conception of constructive or destructive
interference between two hydrogenic parabolic states, resulting in enhanced or
weakened oscillator strength, respectively.

Mathematically, the final state $|f\rangle$ can be written as%
\begin{align}
|f\rangle=c_{1}|red>+c_{2}|blue>
\end{align}
for the simplest case, then the oscillator strength is expressed as
\begin{align}
 c_{i \to f}  &= | < i|d|f > |^2 \nonumber \\
  &= c_1^2 | < i|d|red> |^2  + c_2^2 | < i|d|blue > |^2  \\
  &+ c_1 c_2  < i|d|red>^*  < i|d|blue >  + c.c., \nonumber
\end{align}
where the crossing terms contribute the constructive or destructive interference
dependent on the mixing coefficients $c_1$ and $c_2$.

\begin{figure}[ptb]
\centering
\includegraphics[width=3.3in]{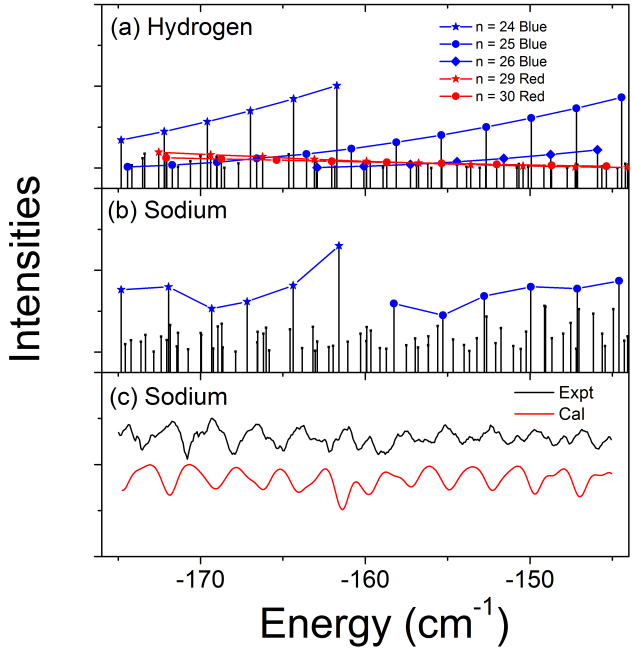}\caption{(Color online) Calculated
oscillator strength above the saddle point $E_{sp} = -177.4
\quad \mathrm{{cm^{-1}}}$ at $F = 840$ V/cm for (a) hydrogen atom, (b) sodium
atom. The calculate spectral lines are convoluted
with a Gaussian broadening and compared with the experimental observations  as well (c).}%
\label{fig5}%
\end{figure}

\begin{table*}[ptb]
\centering {\small \begin{threeparttable}
\caption{The energy level positions (in $\rm{cm^{-1}}$) of the stark resonances of hydrogen and sodium ($F = 840$ V/cm, $m=0$) within $-175$
cm$^{-1}$ $\sim$ $-145$ cm$^{-1}$ corresponding to the transitions in Fig.\ref{fig5}.
}
\begin{tabular}{ccc|cc}
\toprule
\quad & hydrogen&\quad& \quad& sodium \quad \\
\hline
State:$(n_1,n_2,m)$ &\quad Theory$^a$  &\quad Theory$^b$  &\quad Theory$^c$  &\quad Experiment$^d$ \\
\hline
(20,3,0) & -169.589 & -169.589 & -169.319 & -169.320\\
(21,2,0) & -166.972 & -166.971 & -167.172 & -167.054\\
(22,1,0) & -164.352 & -164.351 & -164.378 & -164.113\\
(23,0,0) & -161.729 & -161.729 & -161.599 & -161.348\\
(20,4,0) & -155.388 & -155.388 & -155.289 & -155.289\\
(21,3,0) & -152.655 & -152.655 & -152.778 & -152.366\\
(22,2,0) & -149.920 & -149.920 & -149.954 & -149.718\\
(23,1,0) & -147.181 & -147.181 & -147.131 & -147.036\\
(24,0,0) & -144.440 & -144.440 & -144.592 & -144.592\\
(2,27,0) & -168.273 & -168.735 & -168.710 & -168.710\\
\hline
\end{tabular}
\label{tab2}
\small
\begin{tablenotes}
\item[a] Calculation for hydrogen by the seventh-order perturbation theory \cite{1099};
\item[b] Calculation for hydrogen by the present EQDT method;
\item[c] Calculation for sodium by the present EQDT method;
\item[d] Experimental observation for sodium with total Gaussian broadening
of $0.18$ cm$^{-1}$.
\end{tablenotes}
\end{threeparttable}
 }\end{table*}

\begin{figure*}[ptb]
\centering
\includegraphics[width=5in]{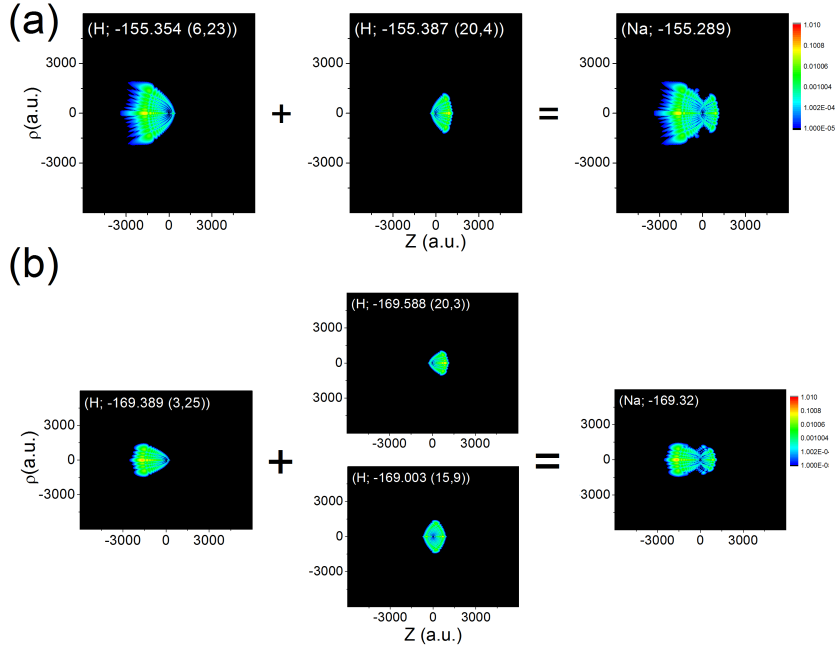}\caption{(Color online) Contour plot of
electron probability distribution $|\Psi(\rho,z)|^{2}$ for some typical states
in Fig. \ref{fig5}. Each state of sodium can be considered as a superposition
of two or more hydrogenic parabolic states.}%
\label{fig6}%
\end{figure*}

For the higher energy levels, for example, above the stark saddle point $E_{c}$,
as shown in Fig.\ref{fig5},
more states come to interact with each other, leading to more complex oscillator
re-distributions.
For hydrogen atom, as can be seen
from Fig.\ref{fig5}(a), the oscillator strength distribution above $E_{c}$ still shows regularity like the states below the saddle point $E_{c}$.
For non-hydrogenic atoms, however,
the
core-induced interaction causes the oscillator strength to behave chaotic completely
as shown in Fig.\ref{fig5}(b) although the broadened experimental spectrum
shows a smooth oscillation in Fig.\ref{fig5}(c).
The oscillation
spacing in our spectrum, $ 2.67$ $\mathrm{{cm^{-1}}}$, is very close to the
estimation by formula  $3e\epsilon{n_{c}}{a_{0}} \simeq(9.16\; \mathrm{{cm^{-1}%
})(\epsilon/4335\;{V/cm})^{3/4}}$ \cite{293}, which gives a value of  $2.68$ $\mathrm{{cm^{-1}}}$ at
electric field of $840$ V/cm.

However, the energy levels above $E_{c}$ have a different feature compared with those below $E_{c}$, that is, the red states above saddle point
have very short lifetime, making the spectral lines extremely broadened.
Like table \ref{tab1}, we also  give the energy level positions for hydrogen and sodium in table \ref{tab2}.
Similar to the case below $E_{c}$, we draw the electron
probability distributions for several typical states given in Fig.\ref{fig5}. They
are shown in Fig.\ref{fig6}.
It's
worth noting that  the red states  above
$E_{c}$ have  extremely short lifetime,  but their coupling with blue states will prolong their
lifetime greatly, providing a chance to   perform `photoionization microscopy'
\cite{1085}.
In Fig.\ref{fig6}(a), the resonant state of sodium at $E=-155.289$ can be
viewed as the superposition of the hydrogen parabolic states $(6,23)$ and $(20,4)$.  Specially, we give a state of sodium with decomposition of one red and two blue
states, which is shown
in Fig.\ref{fig6}(b). Obviously, the blue state $(15,9)$ contributes the
least for the final sodium state at $E=-169.32$. For all the above cases,
however, the components of red or blue states will come to interference close to
the atomic core.

\section{Conclusion}

In summary, we
demonstrate that the electric-field-induced coupling can be clearly visualized
in the form of wave-functions.
The
parabolic separation for hydrogen in electric field can supply a set of
`orthogonal' basis
composed of red and blue states and
the non-hydrogenic behaviour of Rydberg sodium in electric field
can be seen as coupling between hydrogenic red and blue states.
By viewing the
electron probability distributions directly, we show that the non-hydrogenic
wave-function can be
intuitive in understanding the coupling behaviour of non-hydrogenic atom in
static electric fields on benefit of the `orthogonality' of basis. Moreover, the irregular oscillator strength distribution of
sodium is analyzed in the perspective of electric-field-induced interference
between hydrogenic red and blue states. The constructive and destructive interferences
are confirmed
in the wavefunction analysis and well explain the oscillator strength from
theoretical calculation and experimental observation for Rydberg sodium in electric field
of 840 V/cm.

\begin{acknowledgments}
This work was supported by the National Basic Research Program of China under
grant No. 2013CB922003, by the National Natural Science Foundation of China
under grant Nos. 11174329, 91121005 and 91421305.
\end{acknowledgments}

\bibliographystyle{PhysRevA}

\end{document}